\documentclass[reprint, amsmath, amssymb, aps, prb]{revtex4-1}

\usepackage[utf8]{inputenc} % acentuação
\usepackage{braket}
\usepackage{graphicx}% Include figure files
\usepackage{dcolumn}% Align table columns on decimal point
\usepackage{bm}% bold math
\usepackage[normalem]{ulem} % strikethrough
\usepackage{xcolor}

\usepackage{hyperref}% add hypertext capabilities
\hypersetup{
    colorlinks,%
    citecolor=blue,%
    linkcolor=blue,%
    urlcolor=blue
}

\newcommand{\cs}[1]{(#1C$_4$S$_4)_3$}
\newcommand{\fp}{first-principles}

\begin{document}

%\preprint{APS/123-QED}
\title{Quantum anomalous Hall effect in metal-bis(dithiolene), 
\\ magnetic properties, doping and interfacing graphene}

%%\thanks{A footnote to the article title}
%
\author{F. Crasto de Lima}
\author{Gerson J. Ferreira}
\author{R. H. Miwa}
\affiliation{Instituto de F\'isica, Universidade Federal de Uberl\^andia, \\
        C.P. 593, 38400-902, Uberl\^andia, MG,  Brazil}%
\date{\today}

\begin{abstract}

The realization of the Quantum anomalous Hall effect (QAHE) in two dimensional 
(2D) metal organic frameworks (MOFs), \cs{M} with M = Mn, Fe, Co, Ru and Rh, has 
been investigated based on a combination of \fp\ calculations and tight binding 
models. Our results for the magnetic anisotropy energy (MAE) reveal that the  
out-of-plane (in-plane) magnetization is favored for M = Mn, Fe, and Ru (Co, and 
Rh). Given the structural symmetry of \cs{M}, the QAHE takes place only for M = 
Mn, Fe and Ru. Such a quantum anomalous Hall phase has been confirmed through 
the calculation of the Chern number, and examining the formation of 
topologically protected (metallic) edge states. Further electron ($n$-type) 
doping of the MOFs has been done in order to place the Fermi level within the 
non-trivial energy gap; where we find that in \cs{Ru}, in addition to the 
up-shift of the Fermi level, the MAE energy increases  by 40\%. Finally, we show that in MOF/graphene (vdW) interfaces, the Fermi level tunning can be done with an external electric field, which controls the charge transfer at the MOF/graphene interface, giving rise to switchable topologically protected edge currents in MOFs.
\end{abstract}

\maketitle

\section{Introduction}

Since the theoretical proposal of the quantum spin Hall (QSH) phase in 
graphene\,\cite{kanePRL2005}, the search, as well as the control, of the 
topological phases in two dimensional (2D) systems has been the subject of 
intense studies over the past few years. For instance, investigations addressing 
2D materials characterized by larger and tuneable (non-trivial)  energy gaps 
mediated by mechanical strain\,\cite{xuPRL2013}, external electric field 
perpendicular to the 2D sheet (EEF$_\perp$)\,\cite{drummondPRB2012}, and/or 
suitable chemical combinations\,\cite{padilhaSciRep2016}. Further studies have 
been done focusing on the van der Waals (vdW) heterostructures by stacking 2D 
topological insulators (TI), and combinations of trivial/topological 
materials\,\cite{qianSci2014}.

In a seminal work, Wang {\it et al.}\,\cite{wangPRL2013} predicted the quantum 
anomalous Hall effect (QAHE) in a metal organic framework (MOF) composed by 
benzene rings attached to three-fold coordinated Mn atoms. Different topological 
phases have been identified in other  MOFs,  like the the QSH phase in  
nickel-bis-dithiolene [\cs{Ni}]\,\cite{NANO2842Feng}, synthesized by Kambe {\it 
et al.}\,\cite{JACSSakamoto2013};  and  the QAHE in manganese-bis-dithiolene 
[\cs{Mn}]\,\cite{zhaoNanoscale2013}. In the latter, the topological phase has been 
changed from  QSH  to QAH  by replacing the metallic element, Ni$\rightarrow$Mn. 
On the other hand, keeping the transition metal and changing the organic host, 
recent theoretical studies predicted that (MnC$_3$S$_6$)$_3$ can be tuned from a 
Chern insulator to Chern  half-metal as a function of the Fermi 
level\,\cite{wangJPhysChemLett2017}. In parallel, currently we are facing an 
amazing progress on the synthesis of 2D metal 
frameworks\,\cite{LangmuirMaeda2016}.  For instance, the synthesis of  metal  
(M) bis-dicyanobenzenedithiolate with M = Fe, Co, Ni, Pd, Pt, and 
Zn\,\cite{EJICAlves2004}, which is somewhat similar to the metal bis-dithiolene. 
Further molecular design has been done by building up multilayered systems by 
stacking MOFs\,\cite{colsonNatChem2013,sheberlaJACS2014, 
rodriguezChemComm2016,sakamoto2016coordination}. Very recently, based on \fp\, 
calculations, we have found that bilayer systems of \cs{M}, with M=Ni and Pt, present the Z$_2$-metallic phase, where  the edge states can be tuned  by an 
EEF$_\perp$\,\cite{PRBdeLima2017}.

{In contrast with its counterpart, the time-reversal symmetric QSH phase,   the 
breaking of this symmetry by intrinsic magnetization rules the 
emergence of the QAHE.} Such an additional ingredient can be used to control the 
the electronic properties and the topological phases in 2D  systems. Indeed, by 
tuning the strength of {the exchange field}, we may have different types of 
topological gaps and half-metallic phases induced by (intra/inter) 
spin orbital couplings (SOCs)\,\cite{huNanoLett2015,dongPRL2016two}. Meanwhile, 
graphene decorated by 5d transition metals presents a switchable QAHE, where the 
out-of-plane/in-plane  magnetization can be controlled by an 
EEF$_\perp$\,\cite{zhangPRB2012electrically}. Somewhat similar control of the 
topological phase has been predicted for the triphenyl-lead MOF. At the ground 
state it is characterized by a (antiferromagnetic) topologically trivial 
phase, and the QAHE takes place mediated by an EEF$_{\perp}$, which promotes 
an out-of-plane ferrimagnetic phase\,\cite{kimPRB2016}. Very recently, Ren 
{\it et al.} demonstrated that the in-plane magnetization can induce  
the QAHE in 2D systems with inversion symmetry, and no (out-of-plane) mirror 
symmetry\,\cite{PRBRen2016}.

%%%%%FIG
\begin{figure}[h!]
\includegraphics[width=\columnwidth]{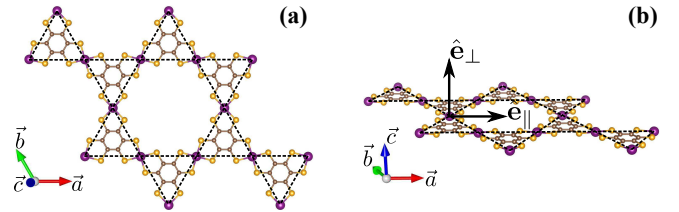}
\caption{\label{str-mae}  Atomic structure of the MOF \cs{M}, top-view (a) and 
side-view (b), the C atoms are shown in brown, S in yellow and M in purple.}
\end{figure}

In this work, we use  \fp\, calculations and tight-binding (TB) models to 
investigate the electronic and magnetic properties of \cs{M} MOFs, with M=Mn, 
Fe, Co, Ru, and Rh. Fig.\,\ref{str-mae} shows the structural model of 
free-standing  \cs{M}. The energetic preference for in-plane versus out-of-plane 
magnetization was examined through the calculation of the magnetic anisotropy 
energy (MAE). We found that the out-of-plane magnetization is favored in \cs{M} 
MOFs with  M=Mn, Fe, and Ru, giving rise to  (intraspin) energy gap induced by 
the SOC. The QAHE of those MOFs was characterized by the calculation of the 
Chern number ($C$), and the formation of topologically protected (metallic) edge 
states. We show that the Fermi level can be tunned in and out of the non-trivial 
energy gap with quite feasible electron ($n$-type) doping. Further MAE 
calculations, on the electron-doped systems, indicate a strengthening of the 
out-of-plane magnetization in \cs{Ru}, namely, the MAE increases from 2.0 to 
2.8\,meV/Ru-atom. Finally, we show that such a $n$-type doping, and 
consequently the  energy position of the Fermi level, can be controlled on 
MOF/graphene vdW interfaces through an EEF$_\perp$.

\section{Method}

     The calculations were performed based on the DFT approach, as implemented 
in the VASP code\cite{vasp1}. The exchange correlation term was described using 
the GGA functional proposed by Perdew, Burke and Ernzerhof (PBE)\cite{PBE}. The 
Kohn-Sham orbitals are expanded in a plane wave basis set with an energy cutoff 
of 400 eV. The 2D Brillouin Zone (BZ) is sampled according to the Monkhorst-Pack 
method\cite{PhysRevB.13.5188}, using a gamma-centered 4$\times$4$\times$1 mesh 
for atomic structure relaxation and 6$\times$6$\times$1 mesh to obtain the 
self-consistent total charge density. The electron-ion interactions are taken 
into account using the Projector Augmented Wave (PAW) method 
\cite{PhysRevB.50.17953}. All geometries have been relaxed until atomic forces 
were lower than $0.025$\,eV/{\AA}. The metal-organic framework (MOF) monolayer 
system is simulated considering a vacuum of $16$\,{\AA} perpendicular to the 
layers. In order to describe the strong Coulomb interaction between $d$ orbitals 
of the transition metal, we adopted the GGA+U approach \cite{PRBDudarev1998}. In 
the formation of interface with graphene the vdW interaction (vdW-DF2 
\cite{PRBLee2010}) was considered to correctly describe the system.

\section{Results and Discussions}

\subsection{Equilibrium geometry and the MAE}

The MOF Metal-Bis(dithiolene)  has a hexagonal atomic structure with the metal 
atoms (M) forming a kagome like structure,  indicated by the dashed lines in 
Fig.\,\ref{str-mae}(a).  Here, we  have considered {M = Mn, Fe, Co, Ru and 
Rh}. The calculated equilibrium lattice constants ($a$), presented in 
Table\,\ref{parameters1}, matches closely those of the (experimentally synthesized) \cs{Ni}, where  $a$ = 14--15\,\AA\,\cite{JACSSakamoto2013}. 

%%%%%%FIG
\begin{figure*}
\includegraphics[width=2\columnwidth]{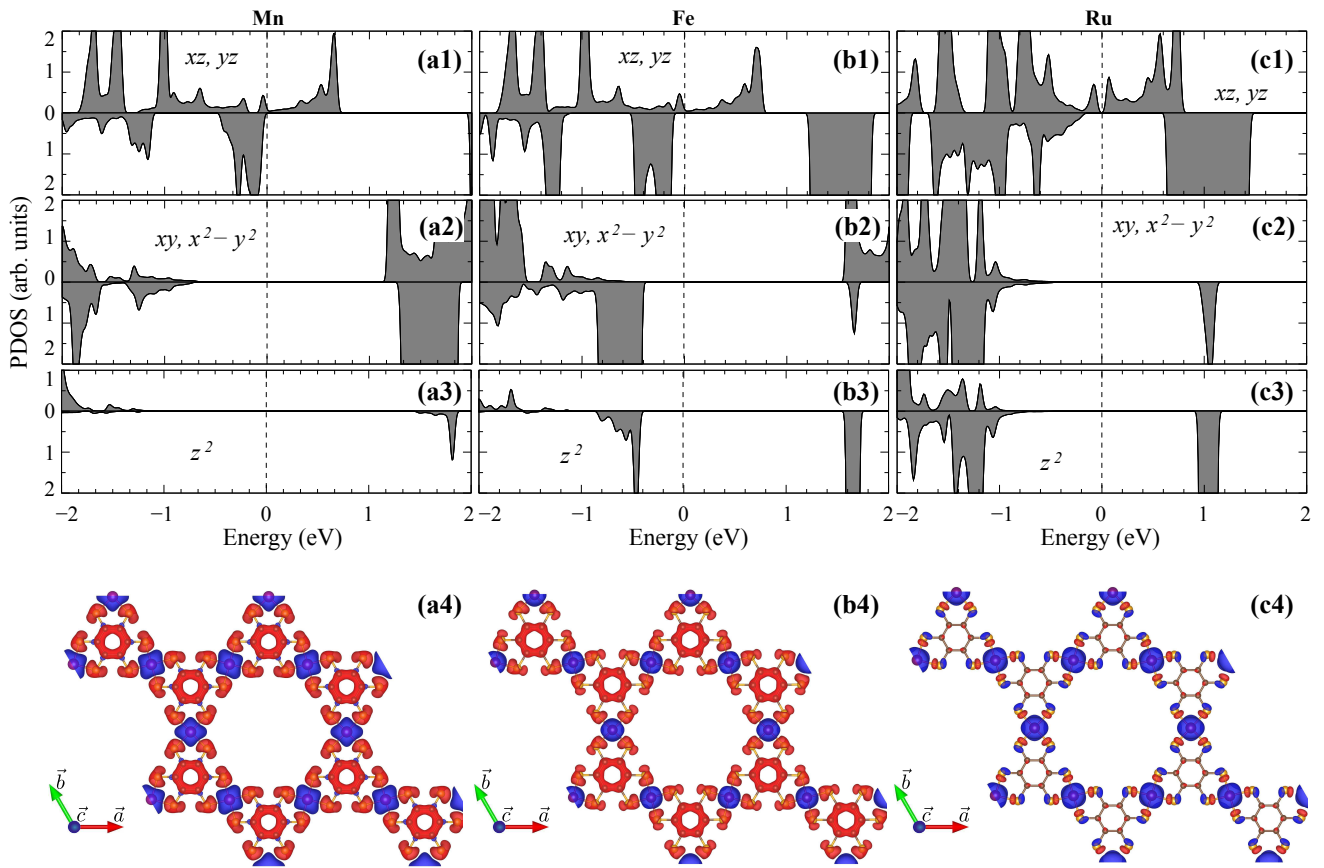}
\caption{\label{pdos} Spin-polarized projected density of states (PDOS) and the 
spin-densities of \cs{Mn} (a1)--(a4), \cs{Fe} (b1)--(b4), and \cs{Ru} 
(c1)--(c4). The spin-density ($\Delta\rho^{\uparrow\downarrow}$) is defined as, 
$\Delta\rho^{\uparrow\downarrow} = \rho_{\uparrow} - \rho_{\downarrow}$, with 
isosurface of $0.002$\,{\AA$^{-3}$} and blue (red) are  regions with net up 
(down) spin density.  }
\end{figure*}	

In the present \cs{M} systems, the QAHE is constrained to an 
out of plane magnetization of the MOF
\cite{PRBRen2016}. The net magnetic moment and its 
orientation in  \cs{M}  are dictated  by  the $3d$ and $4d$  orbitals of the 
transition metals; the preferential magnetic orientation was obtained by 
computing the magnetic anisotropy energy ($E_{\rm MAE}$), here defined as 
$E_{\rm MAE} = E_{\perp} - E_{\parallel}$,  where $E_{\perp}$  and 
$E_{\parallel}$ are the total energies of the MOFs with the magnetic 
moment aligned out-of-plane and in-plane, respectively. It is worth noting 
that the strongly correlated $d$ electrons requires the consideration of the 
Hubbard correction term. Our results, within the GGA+U approach, of magnetic 
anisotropy energy ($E_{\rm MAE}$) and the net magnetic moment ($m$) as a 
function of the Hubbard U values, summarized in Table\,\ref{gga+u} in Appendix A,
reveals that the  stabilization of those magnetic properties is achieved  for 
$\rm U\ge 3$\,eV for all studied metals. 

The obtained values for the $E_{\rm MAE}$ and $m$ are shown in 
Table\,\ref{parameters1} for $\rm U=3$\,eV. We find that  \cs{Mn} presents a net 
magnetic moment of 3.12\,$\mu_{\rm B}$, and easy-axis out-of-plane, $E_{\rm 
MAE}=0.62$\,meV/Mn-atom, which supports the quantum anomalous phase proposed by 
Zhao {\it et al.}\,\cite{zhaoNanoscale2013}. The QAHE has been also predicted in 
graphene adsorbed by transition metals Fe\,\cite{qiaoPRB2010}, Co, 
Rh\,\cite{huNanoLett2015}, and Ru\,\cite{acostaPRB2014}. However, in \cs{Co} and 
\cs{Rh} we found an energetic preference for the in-plane magnetization, $E_{\rm 
 MAE}=-1.84$\,meV/Co-atom and $-2.06$\,meV/Rh-atom, thus indicating that the QAH 
phase is not expected in those MOFs. Meanwhile, \cs{Fe} and \cs{Ru} present  FM 
phases with  $E_{\rm MAE}$=0.31\,meV/Fe-atom and 2.0\,meV/Ru-atom, which points 
to the emergence of QAH phases in those MOFs. 

\begin{table}[h!]
\caption{\label{parameters1} Lattice parameter $a$ ({\AA}); magnetization $m$ 
per TM ($\mu_b$/TM); Magnetic Anisotropy Energy $E_{MAE} = (E_{\parallel} - 
E_{\perp})/N_{M}$ (Kelvin/TM) positive (negative) are for easy axis 
perpendicular (parallel) to the layer; work function $\Phi = V_{vac} - E_f$ 
(eV). All for $U=3.0$\,eV. } \begin{ruledtabular}
\begin{tabular}{c|ccccr}
MOF         &   $a$   &    $m$     & $E_{\rm MAE}$   & $\Phi$ \\
\hline
 \cs{Mn}  &  15.00  &    3.12    &     7.2   &  5.53  \\
 \cs{Fe}  &  14.81  &    2.15    &     3.6   &  5.60  \\
 \cs{Co}  &  14.67  &    1.79    &   -21.4   &  5.75  \\
%\cs{Tc}  &  15.27  &    2.86    &    57.1   &   5.38  \\
 \cs{Ru}  &  15.19  &    1.86    &    23.0   &  5.46  \\
 \cs{Rh}  &  15.07  &    0.86    &   -23.9   &  5.60  \\
\end{tabular}
\end{ruledtabular}
\end{table}

%%%%%%FIG
\begin{figure*}
\includegraphics[width=2\columnwidth]{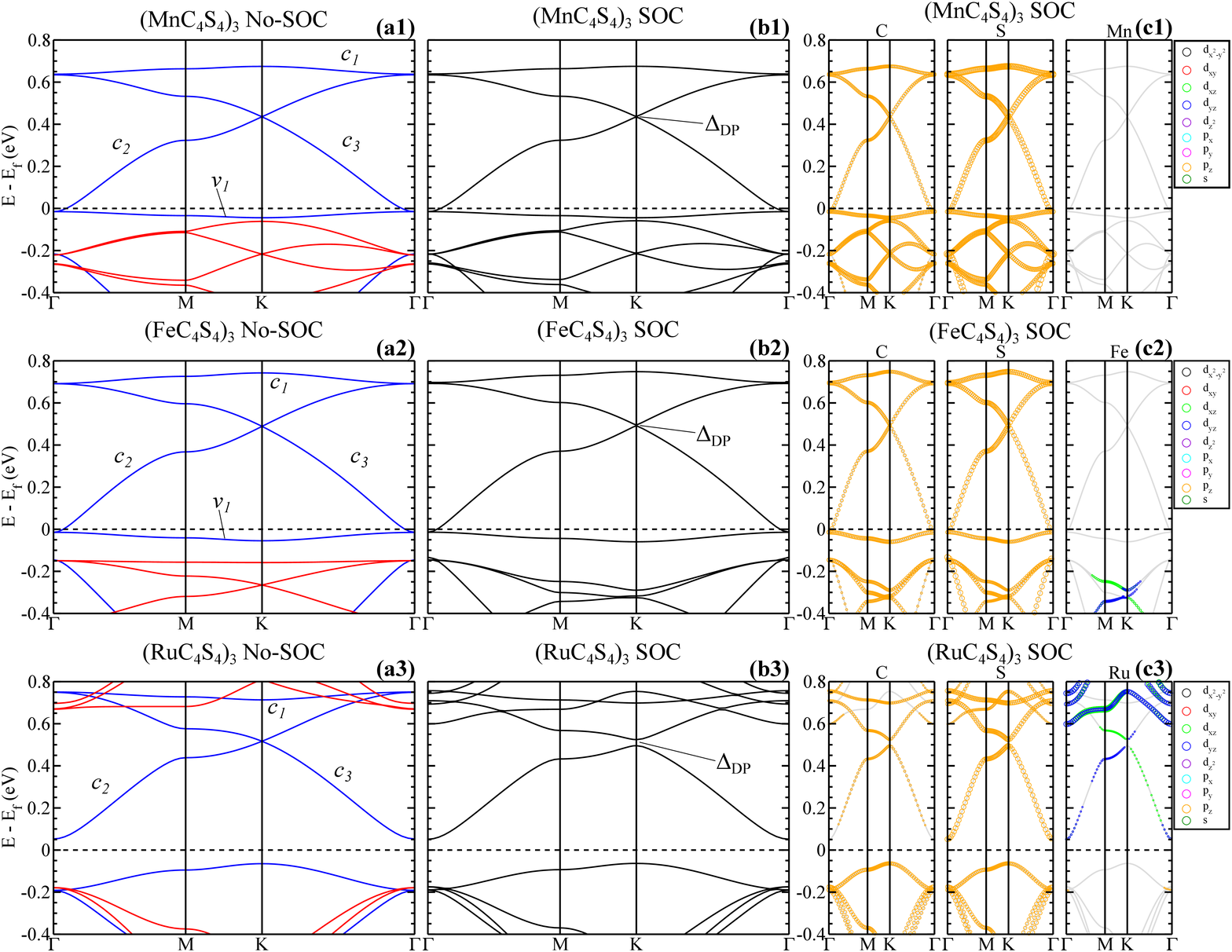}
\caption{\label{bands} Electronic band structures of the MOFs, \cs{Mn}, 
\cs{Fe}, and \cs{Ru} without SOC (a1)--(a3) where blue (red) are for spin up (down) bands, with SOC (b1)--(b3) and the
projected  on the atomic orbitals (c1)--(c3).} 
\end{figure*}	

\subsection{Qualitative analysis of the magnetic anisotropy}

The \cs{M} MOFs present a $D_{6h}$ symmetry, giving rise to  
two sets of doubly degenerated levels at the $\Gamma$ point, $e_{2g}$ (d$_{x^2-y^2}$ 
and d$_{xy}$ orbitals), and $e_{1g}$ (d$_{xz}$ 
and d$_{yz}$ orbitals), and a non-degenerated level $a_{1g}$ ($d_{z^2}$ 
orbital). The projected density of states (PDOS) of these transition metal $d$ orbitals, calculated without SOC, are shown in Fig.\,\ref{pdos}.

 The net magnetization of the MOFs is (i) mostly dictated by the 
$d$ orbitals of the the transition metals, where  about 50\% of the magnetic 
moment comes from the unpaired $d_{xz,yz}$ orbitals, while  the out of plane 
$d_{z^2}$ orbitals contribute with  40\% in \cs{Ru}, followed by \cs{Fe}  and 
\cs{Mn}, 30\% and 20\%, respectively. As a consequence, (ii) the in plane 
polarization is larger (lower) in the latter (former) system.  Those transition 
metals are embedded in an organic  framework, composed  C and S atoms, which 
present opposite net magnetization ruled by the C-2$p$ and S-3$p$ orbitals;  
such a (opposite) spin polarization is larger (lower) in \cs{Mn} (\cs{Ru}). Our  
spin density results, $\Delta\rho^{\uparrow\downarrow} = \rho_{\uparrow} - 
\rho_{\downarrow}$ [Figs.\,\ref{pdos}(a4)--(c4)], provide a picture of (i) and 
(ii) discussed above, and allow us to infer that the FM coupling between the 
transition metals is given by an indirect exchange process, mediated by the 
organic host of C and S atoms.

The perturbation approach proposed by Wang {\it et al.}\,\cite{wangPRB1993}, 
allow us to use the PDOS above to provide a qualitative analysis of our MAE 
results\cite{songPRMat2017}. There, the $E_{\rm MAE}$ is approximately defined 
by the matrix elements of the orbital angular momentum, {\it i.e.} $\langle 
u_\sigma|L_z|o_{\sigma'}\rangle$ and $\langle u_\sigma|L_x|o_{\sigma'}\rangle$, 
where $\ket{o_\sigma}$ and $\ket{u_\sigma}$ refer to the orbital component of 
the wave-function (e.g. $\ket{\psi} = \ket{u_\sigma}\otimes\ket{\sigma}$) of the 
occupied ($o$) and empty ($u$) states with a given spin $\sigma = \pm$, 
calculated without SOC. Namely, the magnetic anisotropy energy can be written as 
$E_{\rm MAE} = \Delta E_{\rm d} + \Delta E_{\rm nd} $, where the spin diagonal 
and non-diagonal terms are

\begin{align}
	\Delta E_{\rm d} &= \xi^2 \sum_{\sigma, o,u} \dfrac{|\bra{u_\sigma} L_z \ket{o_\sigma}|^2 - |\bra{u_\sigma} L_x \ket{o_\sigma}|^2}{\varepsilon_{u,\sigma}- \varepsilon_{o,\sigma}}, \label{eq:EmaeD}
	\\
	\Delta E_{\rm nd} &= \xi^2 \sum_{\sigma, o,u} \dfrac{|\bra{u_\sigma} L_x \ket{o_{\bar{\sigma}}}|^2 - |\bra{u_\sigma} L_z \ket{o_{\bar{\sigma}}}|^2}{\varepsilon_{u,\sigma}- \varepsilon_{o,\bar{\sigma}}}, \label{eq:EmaeND}
\end{align}
where $\xi$ represents the strength of the SOC. The matrix elements are divided by their (respective) single particle energy difference, indicating that the dominant contributions arise from states near the Fermi level. 

For $M = \text{Mn, Fe, and Ru}$, this model tell us that the $E_{\rm MAE}$ is 
defined by the $e_{1g}$ (d$_{xz}$ and d$_{yz}$ orbitals) orbitals, due to the 
PDOS peaks near the Fermi level in Figs.\,\ref{pdos}(a1)--(c1). In these cases, 
the empty states near the Fermi level are all of the $\ket{u_+}$ type. Moreover, 
group theory selection rules dictates that $L_x$ matrix operators between 
$e_{1g}$ states are null. Therefore the dominant contributions are given by the 
terms $|\bra{u_+} L_z \ket{o_+}|^2 - |\bra{u_+} L_z \ket{o_-}|^2$ in 
Eqs.~\eqref{eq:EmaeD}-\eqref{eq:EmaeND}. That is, the sign of $E_{\rm MAE}$ is 
given by the competition between the positive contribution from the 
spin-diagonal $L_z$ matrix element, and the negative contribution from the 
spin-flipped $L_z$ terms. For Ru, Fig.~\ref{pdos}(c1), there are no $\ket{o_-}$ 
contributions near the Fermi level to the PDOS, yielding a large $E_{\rm MAE}$ 
due to the positive spin-diagonal term. For Fe and Mn, 
Figs.~\ref{pdos}(a1)--(b1), the $\ket{o_-}$ PDOS peaks approach the Fermi level, 
reducing the $E_{\rm MAE}$.

%%%%%%FIG
\begin{figure}[h!]
\includegraphics[width=\columnwidth]{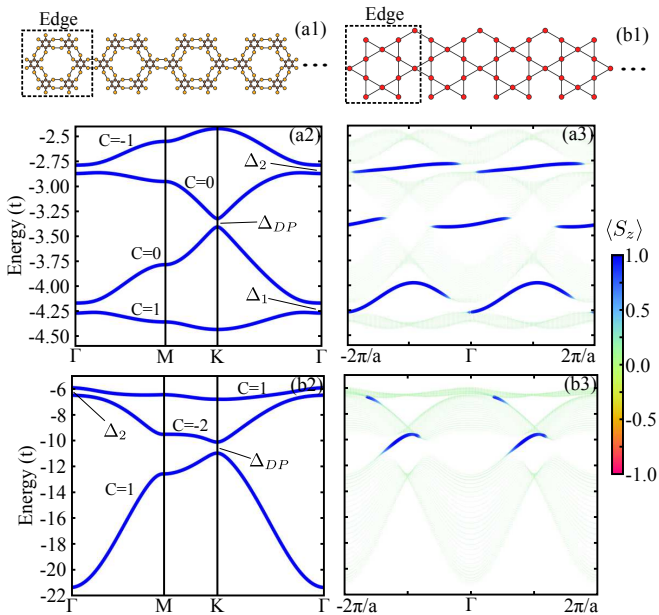}
\caption{\label{tb} Lattice structure for the TB model showing the edge geometry 
for the C-S lattice (a1) and the kagome lattice (b1). 2D bulk band structure for 
C-S (a2), kagome (b2), and band structure projected edges states  for C-S (a3) 
and kagome (b3). The color bar indicates the spin projection for the bulk and NR band structure.}
\end{figure}

\subsection{Kagome bands}

The  exchange field induced by  the transition metals gives rise to 
(spin-polarized) kagome bands. As shown in
Figs.\,\ref{bands}(a1)-(c3), those energy bands  lie within an 
energy interval of about 0.8\,eV above the Fermi level; characterized by nearly flat  bands ($c_1$ and $v_1$) degenerated with Dirac like energy 
bands, $c_2$ and $c_3$. Those energy bands ($c_2$ and $c_3$) present linear 
energy dispersion at the K-point, Dirac point (DP).  The electronic band 
structures  of \cs{Mn} and \cs{Fe} MOFs reveal a  half-metal behavior, 
characterized by spin-up metallic bands $c_2$ and $c_3$ (partially occupied near 
the $\Gamma$-point),  and semiconductor  spin-down energy bands. Indeed, such a 
half-metal character in \cs{Mn} was already reported by Zhao {\it et 
al.}\,\cite{zhaoNanoscale2013}. 

By turning-on the spin-orbit coupling (SOC),  we find non-trivial energy 
gaps of 2 and 3\,meV at the Dirac point in \cs{Mn} and \cs{Fe}, respectively 
[$\Delta_{\rm DP}$ in Figs.\,\ref{bands}(b1) and (b2)].  For both systems, the 
projection on  the atomic orbitals  [Figs.\,\ref{bands}(c1)  and (c2)] show that 
the kagome bands ($c_1$--$c_3$ and $v_1$), are mostly composed by C- and S-$p_z$ 
orbitals, with no contributions from the Mn and Fe transition metals. In 
contrast, as shown in Fig.\,\ref{bands}(a3), \cs{Ru} exhibits a semiconducting 
character for both spin channels. Here the spin-up unoccupied bands form a 
kagome set of bands ($c_1$--$c_3$), with a flat band ($c_3$) degenerated with 
two Dirac like bands ($c_2$, $c_3$). At 0.52\,eV above the Fermi level, the  
degeneracy of the Dirac point has been removed by the SOC, giving rise to a 
non-trivial energy gap of 28\,meV, $\Delta_{\rm DP}$ in  
Fig.\,\ref{bands}(b3). Such a larger  energy gap, compared with the other MOFs, 
can be attributed to the hybridization of the Ru-4$d_{xz,yz}$ with the host C- 
and S-$p_z$ orbitals near the DP of the kagome bands, Fig.\,\ref{bands}(c3).

\subsection{Quantum Anomalous Hall Effect}

In order to provide further support and a more clear physical picture of the   
QAHE in the MOFs, we calculate the Chern number of \cs{M}, and examined the 
formation of topologically protected metallic edge states in \cs{M} nanoribbons. 
The calculations were performed within the TB model by considering one orbital 
per site. Here, based on the projected energy bands [Fig.\,\ref{bands}], we have 
considered  (i) four Hexagonal Multi-Orbital (HMO) bands, composed by the carbon 
and sulfur (C-S) $p_z$  lattice [Fig.\,\ref{tb}(a1)], for the \cs{Mn} and 
\cs{Fe} MOFs; and (ii)  Ru kagome lattice model [Fig.\,\ref{tb}(b1)] for  
\cs{Ru}. The TB Hamiltonian can be written as\,\cite{PRLTang2011, 
PRBdeLima2017},

\begin{equation} 
H_{TB} = H_0 + H_{SO} + H_{M} 
\end{equation}
where $H_0$ describe the on site energy and nearest and next-nearest neighbor hopping,
\begin{equation}
H_{0} = \sum_i \varepsilon_i c_i^\dagger c_i + \sum_{i,j} t_{ij} \, c_{i}^{\dagger} c_{j},
\end{equation}
$H_{SO}$ is the intrinsic spin-orbit coupling of the sites,
\begin{equation}
H_{SO} = i\,\sum_{i,j} \lambda_{ij} \, c_i^{\dagger} {\bm \sigma} \cdot \frac{\left( {\bm d}_{kj} \times {\bm d}_{ik}\right)}{|d_{kj}||d_{ik}|}c_j,
\end{equation}
and $H_M$ is a Zeeman exchange field,
\begin{equation}
H_M = b \sum_{i} c_i^{\dagger} \sigma_z c_i.
\end{equation}
Here $c_{i}^{\dagger} = (c_{i\uparrow}^\dagger,\, c_{i\downarrow}^\dagger )^T$ 
and $c_{i} = (c_{i\uparrow},\, c_{i\downarrow} )^T$, with $c_{i\alpha}^\dagger$ 
and $c_{i\alpha}$ are the creation and annihilation operators of an electron in 
the $i$-th site with spin $\alpha$; $\bm{\sigma}$ stands for the Pauli matrices 
in the spin space. The ${\bm d}_{ij}$ are the vectors connecting the $i$-th to 
the $j$-th site, $\varepsilon_i$, $t_{ij}$,$\lambda_{ij}$ and $b$ controls the 
on-site, hopping, spin-orbit and Zeeman field strength, respectively. The 
hopping and spin-orbit terms depend on the distance between the sites, 
respectively
	\begin{eqnarray*}
	t_{ij} &=& -N t \exp \left\{ - \alpha \, d_{ij} \right\},	\\
	\lambda_{ij} &=& N \lambda \exp \left\{ - \beta \, d_{ij} \right\}.
	\end{eqnarray*}
We have taken the normalization with respect to the first neighbor distance 
$d_{nn}$, $N=\exp(-\alpha d_{nn})$, and $t=1$; $\alpha$ and $\beta$ 
control of the range of the hopping, namely, for larger values ($\gg 1$)  only 
first neighbor hopping has been considered, while for $\alpha$, $\beta \ll 1$, 
further neighbors are taken into account. In Figs.\,\ref{tb}(a2)--(b2), we 
present the the energy bands  for the C-S lattice, {\it i.e.}  \cs{Fe} and 
\cs{Mn} 
[(i)], and the Ru kagome lattice [(ii)], where we find that the  main features 
of the DFT bands have been  nicely captured by the current TB 
approach\,\cite{tb-param}.

Based on the same TB approach, the topological character of the energy 
gaps was  evaluated through the calculation of the Chern number ($C$). By using 
the C-S model, as depicted in Fig.\,\ref{tb}(a2), we found $C = 1$ for the 
Fermi level lying in the energy gap $\Delta_1$, keeping the same values for the 
energy gaps $\Delta_{\rm DP}$ and $\Delta_2$. Meanwhile, based on  the Ru 
kagome lattice model, the    Chern number is positive ($C=1$) at $\Delta_{\rm 
DP}$,  and becomes negative ($C=-1$) in $\Delta_2$, Fig.\,\ref{tb}(b2). It is 
worth noting that the latter is a local gap at the $\Gamma$ point, since the 
energy bands become metallic due to the downward dispersion of the nearly flat 
kagome band. Since the anomalous Hall conductance is given by 
$\sigma_{xy}=(e^2/h)C$, it is predicted  just one conducting channel per edge 
for a given energy position of the Fermi level, where opposite edges present 
conducting channels with opposite carrier velocities 
($\nabla\varepsilon(\bf{k})/\hbar = \bf{v}({\bf k})$).

Indeed, this is what we found along the edge sites of  \cs{M} of the NRs. We 
have considered NR structures with a width of 30 unit cells ($\approx 
44$,\,{nm}); and  based on the  C-S lattice model, in  Fig.\,\ref{tb}(a3) we 
present the calculated electronic band structure projected on the edge sites of 
the NR [dashed-square region  in the Fig.\,\ref{tb}(a1)]. We find only one 
conducting channel, {\it i.e.} only one  topologically protected edge state 
associated with each non-trivial energy gap. All of them are characterized by 
the same direction for the carriers velocities, for instance $\vec b$ in 
Fig.\,\ref{pdos}, and positive values of $\langle S_z \rangle$. Whereas, 
the  opposite edge of the NR will present carrier velocities in the opposite 
direction ($-\vec b$). Similarly, we find the emergence of topologically 
protected edge states in \cs{Ru} NRs described by the Ru kagome 
lattice model, Fig.\,\ref{tb}(b1). As depicted in  Fig.\,\ref{tb}(b3), 
those edge states present one single conducting  channel with  
positive values of  spin-polarization, and  carrier velocities  one along 
the $\vec b$ direction, and  another  one (characterized by a low density 
of states) with carrier velocities in the opposite direction, $-\vec b$. Here, 
the former is associated with the non-trivial energy gap $\Delta_{\rm DP}$ with 
$C=1$, while the latter is due to the local non-trivial energy gap $\Delta _2$ 
with $C=-1$.

\subsection{MOF/G Charge transfer and the  control the Fermi level}

The manifestation of the QAHE, {\it i.e.} the appearance of (topologically) 
protected edge currents, relies on the energy position of the Fermi level 
with respect to the non-trivial energy gaps. In the present  MOFs, as indicated 
in Fig.\,\ref{bands}, we may place the Fermi level in the energy gap 
$\Delta_{\rm DP}$ through electron doping processes. Indeed, this is what we 
obtained by charging the MOF by about $5\times 10^{13}$\,$e$/cm$^2$, which is a 
quite feasible $n$-type doping. It is worth noting that the net magnetization of 
the MOFs increases by about the same amount 0.2\,$\mu_{\rm B}$, while the MAEs 
of the  charged \cs{Mn} and \cs{Fe} MOFs are practically the same when compared 
with those  of the neutral MOFs; whereas in \cs{Ru}, $E_{\rm MAE}=2.0\rightarrow 
2.8$\,meV/Ru-atom.  The required doping to achieve the QAH state in the \cs{Ru} 
makes the magnetic properties of the system more robust with respect to thermal 
perturbations. Such a $n$-type doping can be done through the physisorption of 
the MOF on metal surfaces with work function smaller compared with that of the 
MOF; or we may consider an interface engineering based on 2D 
vdW heterostructures like MOF/graphene.

%%%%%%FIG
\begin{figure}[h!]
\includegraphics[width=\columnwidth]{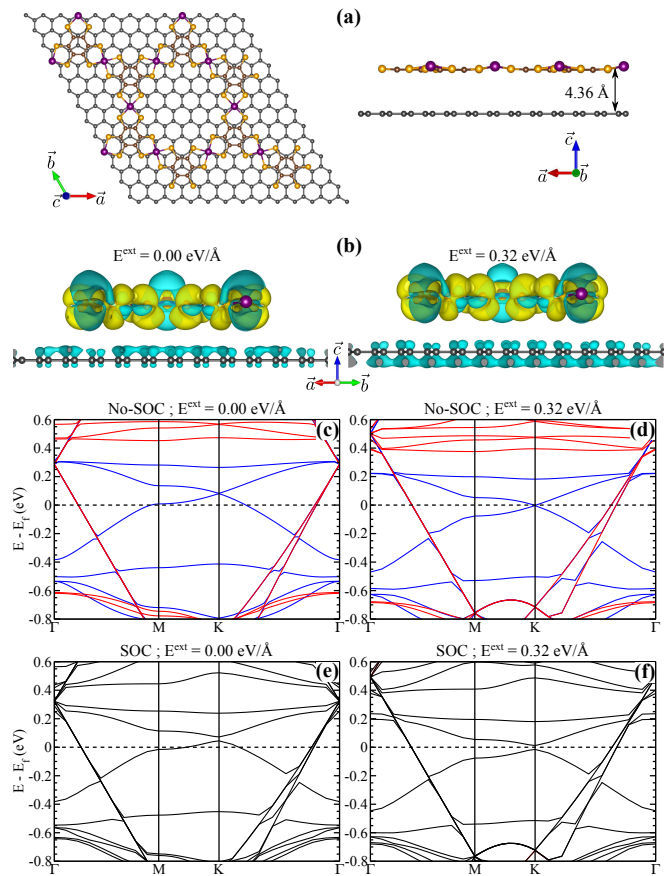}
\caption{\label{ru-grf} Atomic structure of \cs{Ru}/Graphene, top-view and 
side-view  (a), and the charge transfer without and with external electric 
field (b), blue (yellow) indicates regions with loss (gain) of electrons. 
Electronic band structure without SOC and with external electric field 
($E^{ext}$) of $0.00$ (c), $0.32$\,eV/{\AA} (d) and with SOC and $E^{ext}=0.00$ 
(e), $=0.32$\,eV/{\AA} (f). }
\end{figure}

Graphene is a quite interesting substrate, since it provides both structural 
stability, and the $n$-type doping of the MOFs discussed above. The  work 
functions   ($\Phi$) of the MOFs (see Table\,\ref{parameters1}) are about 1\,eV 
larger compared with the that of graphene. Here we will focus on a particular 
MOF/graphene interface,  \cs{Ru}/G\,\cite{mismatch}. 

First, by considering a  set of \cs{Ru}/G staking geometries, we find that the 
energetically most stable configuration is characterized by Ru atoms aligned on 
top of C atoms of the graphene sheet, as seen in Fig.~\ref{ru-grf}(a). The 
energetic stability of \cs{Ru}/G was inferred by the calculation of the binding 
energy $E^b=8.5$\,meV/\AA$^2$, which is defined as the total energy difference 
between the final system [\cs{Ru}/G] and the sum of the total energies of the 
isolated components, \cs{Ru} and single layer graphene. It is worth noting that 
this binding is larger than the one obtained for a graphene bilayer system, 
2.2\,meV/\AA$^2$. At the equilibrium geometry we find an interlayer distance of 
$4.36$\,\AA, Fig.\,\ref{ru-grf}(a), with no chemical bonds at the \cs{Ru}/G 
interface. 

The larger work function of \cs{Ru} ($\Phi=5.46$\,eV), compared with the one  of 
graphene, will promote a net charge transfer from graphene to the MOF. In 
Fig.\,\ref{ru-grf}(b) we present a map of such a charge transfer, where the blue 
(yellow) regions illustrate the loss (gain) of electrons in each layer. 
Focusing on the electronic properties, based on the first-principles 
calculations, upon the formation of the \cs{Ru}/G interface, we find an 
up-shift of the Fermi level of \cs{Ru},  characterizing the $n$-type doping of the 
MOF, whereas the graphene sheet becomes $p$-type doped, with the Fermi level 
lying at 0.3\,eV below the Dirac point. By turning on the SOC, 
Fig.\ref{ru-grf}(e), the non-trivial energy gap $\Delta_{\rm DP}$ of the MOF is 
the same  compared with the one of isolated system; however,  we find that the 
Fermi level of the \cs{Ru}/G system does not lie within $\Delta_{\rm DP}$, and 
thus, there is no appearance  of anomalous Hall current along the edge sites.

In a previous study\cite{PRBdeLima2017}, we verified that the occupation of the  
kagome bands,  as well as the localization of the non-trivial energy gaps in 
MOF/MOF interfaces, can be controlled by an external electric field. It is worth 
to pointing out that the energy bands responsible to the QAH phase are 
completely spin polarized and separated, with the magnetization out of the MOF 
plane. In such a situation, the Rashba SOC does not affect the gap opened by the 
intrinsic SOC, thus contrasting with its non-magnetic QSH 
counterpart\cite{PRBdeLima2017}. Here, mediated by the MOF$\leftrightarrow$G 
charge transfers, the energy bands of the MOF and G system can be tuned by an 
EEF$_\perp$ (perpendicular to the MOF/G interface). In particular, by increasing 
the $n$-type doping of \cs{Ru} by an EEF$_\perp$ of 0.32\,eV/\AA, we find the 
Fermi level lying on the  DP of the MOF [Fig.\,\ref{ru-grf}(d)]; while the 
 SOC gives rise to $\Delta_{\rm DP}$ of 27\,meV at the Fermi level, 
Fig.\,\ref{ru-grf}(f), promoting the (topologically) protected edge currents in  
\cs{Ru}/G, switchable on-and-off through an EEF$_\perp$.

\section{Summary and Conclusion}

Based on combined \fp\, DFT calculations and the tight-binding models, we have 
studied the electronic and the topological  properties of \cs{M} MOFs, with M = 
Mn, Fe, Co, Ru, and Rh. Our results of magnetic anisotropy energy (MAE)  reveal 
that the \cs{M} systems  (with M = Mn, Fe and Ru)  present  the easy axis 
out-of-plane, while the other metals (Co and Rh) are characterized by  an 
energetic preference for in-plane magnetization. Since both inversion and 
out-of-plane mirror reflection symmetries are preserved, the  QAH phase is not 
expected in the latter group\,\cite{PRBRen2016}. Meanwhile, the QAHE of the 
former MOF group was verified through the calculation of the (non-zero) Chern 
number, and  the formation of the topologically protected edge states in MOF 
nanoribbons. Among the currently studied systems, \cs{Ru} presents the largest 
energy gap induced by the SOC. Upon $n$-type doping ($5\times10^{13}e$/cm$^2$), 
in order to place the Fermi level  in the non-trivial energy gap, we find that 
the MAE of \cs{Ru} increases by about 40\%, thus improving its magnetic 
stability with respect to  thermal perturbations. Such a tuning of the Fermi 
level, and the strengthening  the out-of-plane magnetization in \cs{Ru} 
have been examined by  considering the MOF adsorbed on the graphene 
sheet. The formation of the \cs{Ru}/G interface is an exothermic process, which 
may promote the structural stability of the MOF. Further tuning of the Fermi 
level  has been done through  an external electric field, EEF$_\perp$, which 
control the   G\,$\rightarrow$\,\cs{Ru} net charge transfer, and the appearance 
of topologically protected (spin-polarized) edge currents in the MOF. Here, 
although we have considered a single system, \cs{Ru}/G, we may infer that other 
MOF/G interfaces can be engineered to exploit/control the QAH properties, 
like magnetic anisotropy energy and the energy position of the Fermi 
level.

\section{ACKNOWLEDGMENTS}

The authors acknowledge financial support from the Brazilian agencies CNPq, and 
FAPEMIG, and the CENAPAD-SP and Laboratório Nacional de Computação Científica 
(LNCC-SCAFMat) for computer time.

\bibliography{bib}% Produces the bibliography via BibTeX.

\begin{thebibliography}{35}
\bibitem [{tb-()}]{tb-param}%
  \BibitemOpen
  \href@noop {} {}\bibinfo {note} {For the (C--S) hexagonal multi-orbital [(i)]
  we have considered the following parameterization, $a=14.8$\,{\AA},
  $\varepsilon_C=0$\,$t$, $\varepsilon_S=0.5$\,$t$, $\lambda=0.1$\,$t$,
  $b=-1.5$\,$t$ and $\alpha = \beta = 0.6$\,{\AA$^{-1}$}; and for the Ru kagome
  lattice [(ii)] we have used, $\varepsilon=0$\,$t$, $\lambda=0.4$\,$t$,
  $b=-5.0$\,$t$ and $\alpha = \beta = 0.3$\,{\AA$^{-1}$}.}\BibitemShut {Stop}%
\bibitem [{mis()}]{mismatch}%
  \BibitemOpen
  \href@noop {} {}\bibinfo {note} {In order to satisfy the periodic boundary
  conditions, within our supecell approach, the graphene sheet was described by
  using a supercell with periodicity of (6$\times$6), giving rise to lattice
  mismatch of 2.9\%}\BibitemShut {NoStop}%
\end{thebibliography}

\newpage

\appendix

\section{GGA+U}

In Table\,\ref{gga+u} we summarize our results of magnetic anisotropy energy 
($E_{\rm MAE}=(E_{\parallel} - E_{\perp})/N_{\rm M}$) of \cs{M} for M=Mn, Fe, 
Co, Ru, and Rh. {Qualitatively, our results do not depend on the value of $U$, 
which is only required for the quantitative description of the metal $d$ 
orbitals.}

\begin{table}[h!]
\caption{\label{gga+u} Magnetic Anisotropy Energy $E_{\rm MAE}$ (in meV/M-atom) 
and the  magnetization ($m$) of the unity cell per transition metal (M), for 
\cs{M} (M$=$ Mn, Fe, Co, Tc, Ru, Rh), as a function of $U$ ($U=2-4$), positive 
(negative) are for easy axis $\hat{\bf e}_{\perp}$,  perpendicular ($\hat{\bf 
e}_{\parallel}$, parallel) to the layer.} 
\begin{ruledtabular}
\begin{tabular}{crcrcrc} 
\multicolumn{1}{c}{$U$} & 
\multicolumn{2}{c}{2\,eV}   &
\multicolumn{2}{c}{3\,eV}   &
\multicolumn{2}{c}{4\,eV}   \\
\hline
 MOF      & MAE    & $m$ & MAE   & $m$ & MAE   & $m$  \\  
\cline{1-1}
\cline{2-3}
\cline{4-5}
\cline{6-7}
\cs{Mn}   &  0.63 & 3.06 &  0.62 & 3.12 &  0.61 & 3.18   \\
\cs{Fe}   &  0.28 & 2.90 &  0.31 & 2.15 &  0.39 & 2.22   \\
\cs{Co}   & -1.61 & 1.11 & -1.84 & 1.17 & -2.08 & 1.23   \\ 
\cs{Ru}   &  2.14 & 1.78 &  1.98 & 1.86 &  1.84 & 1.92   \\
\cs{Rh}   & -1.29 & 0.79 & -2.07 & 0.86 & -3.00 & 0.90   \\
\end{tabular}
\end{ruledtabular}
\end{table}

\end{document}